\documentclass[draft]{agujournal2019}
\usepackage{url} 
\usepackage{lineno}
\usepackage[inline]{trackchanges} 
\usepackage{soul}

\draftfalse

\journalname{Geophysical Research Letters}

\begin{document}


\title{Seasonal Variability of the Daytime and Nighttime Atmospheric Turbulence Experienced by InSight on Mars}

%
%

\authors{A. Chatain\affil{1}, A. Spiga\affil{1,2}, D. Banfield\affil{3}, F. Forget\affil{1}, N. Murdoch\affil{4}}

\affiliation{1}{Laboratoire de Météorologie Dynamique/Institut Pierre-Simon Laplace (LMD/IPSL), Centre National de la Recherche Scientifique (CNRS), Sorbonne Université, Paris, France}
\affiliation{2}{Institut Universitaire de France (IUF), Paris, France}
\affiliation{3}{Cornell Center for Astrophysics and Planetary Science, Cornell University, Ithaca, NY, USA}
\affiliation{4}{Institut Supérieur de l'Aéronautique et de l'Espace (ISAE-SUPAERO), Toulouse, France}

\correspondingauthor{Audrey Chatain}{audrey.chatain@lmd.ipsl.fr}

\begin{keypoints}
\item InSight's pressure sensor unveils seasonal variability of daytime and nighttime atmospheric turbulence on Mars
\item Local turbulence ($>0.01$~Hz) is sensitive to wind at all local times and seasons contrary to non-local turbulence ($<0.01$~Hz, ie plumes/cells)
\item Northern autumn/winter hosts a remarkable burst of daytime vortices, and nighttime turbulence (including vortices) triggered by strong wind
\end{keypoints}

%
%

\begin{abstract}

The InSight mission, featuring continuous high-frequency high-sensitivity pressure measurements, is in ideal position to study the active atmospheric turbulence of Mars. 
Data acquired during 1.25 Martian year allows us to study the seasonal evolution of turbulence and its diurnal cycle.
We investigate vortices (abrupt pressure drops), local turbulence (frequency range~$0.01-2$~Hz) and non-local turbulence often caused by convection cells and plumes (frequency range~$0.002-0.01$~Hz). 

Contrary to non-local turbulence, local turbulence
is strongly sensitive at all local times and seasons to the ambient wind. 
We report many remarkable events with the arrival of northern autumn at the InSight landing site: a spectacular burst of daytime vortices, the 
appearance of nighttime vortices, and the development of nighttime local turbulence as intense as its daytime counterpart.
Nighttime turbulence at this dusty season appears as a result of the combination of 
a stronger low-level jet, producing shear-driven turbulence, and a weaker stability.

\end{abstract}

\section*{Plain Language Summary}

The weather station on board the InSight lander on Mars includes a very sensitive barometer to measure atmospheric pressure all the time.
We use pressure records by InSight during more than a Martian year to study how the fast (from seconds to minutes)
changes in the atmosphere (named turbulence) varies with seasons on Mars.

Because of the heating of Mars surface by sunlight, turbulence during the day is strong and takes the form of convection plumes and whirlwinds that are detected by InSight.
At night, turbulence is usually not expected because the colder surface prevents convection to occur.
We discovered that turbulence at InSight landing site was unusual in autumn/winter: 
daytime whirlwinds are much more numerous, whirlwinds form even during the night, and there is nearly as much turbulence 
during the night than during the day. 
We explain the increase of nocturnal turbulence at this dusty season on Mars by warmer and windier nights.

%
%


\section{Introduction}
Mars is prone to atmospheric turbulence. Its thin, nearly cloudless atmosphere and low surface thermal inertia entail strong near-surface unstable temperature gradients, resulting in convective turbulence -- cells, plumes, vortices -- at daytime in the Planetary Boundary Layer (PBL)
\cite{Till:94,Lars:02,Petr:11}. 
This induces a wealth of daytime high-frequency variations in the atmospheric pressure, wind and temperature measured by landers and rovers (\citeA{Mart:17} for a review).
Conversely, nighttime conditions in Mars' PBL are highly stable because of strong radiative cooling efficiently inhibiting convection.
Nevertheless, in the night, shear-driven weak turbulence may exist on Mars (e.g., \citeA{Savi:99,PlaG:20}) and, intriguingly, vortex-induced pressure drops have been observed 
in Gale Crater by Curiosity \cite{Ordo:18,Ordo:20}. 

The NASA InSight mission \cite{Bane:19nat} landed to 4.5°N, 135°E in equatorial Elysium Planitia in November 2018. 
The InSight lander operates an unprecedented geophysical station on Mars,
notably permitting seismic measurements \cite{Bane:19nat,Logn:20nat,Giar:20nat},
and features state-of-the-art meteorological sensors \cite{Banf:18,Spig:18insight}.
Atmospheric measurements of the first 400 sols of the mission revealed a sustained daytime vortex activity, quasi-periodic signals associated with convective cells, nighttime gravity waves, 
and a quiet regime 2 to 4 hours after sunset \cite{Banf:20,Spig:21jgr,Lore:20catalog}.
This quiet regime follows the collapse of the daytime convective PBL before the nighttime development of the nocturnal low-level jet \cite{Savi:93,Josh:97llj},
during which weaker turbulence than in the daytime is experienced by InSight sensors \cite{Garc:20,Char:21}.
Apart from its interest for Mars' weather, studying PBL turbulence is key to characterize atmosphere-induced seismic noise \cite{Stut:21,Ceyl:21}.

The aim of this paper is to carry out a study of the diurnal and seasonal variability of atmospheric turbulence monitored by the InSight mission over one Martian year (plus one season, a total of 837 observed sols). 
Following e.g. \citeA{Cola:13}, we distinguish
turbulent structures as vortices, local turbulence (smaller-scale and high-frequency fluctuations above 0.01~Hz, excluding vortices) and non-local turbulence (larger, structured, turbulent features -- mainly convection plumes and cells during the day -- which have longer timescales up to about a quarter of an hour, the convective overturning timescale).
We report here many unexpected events with the arrival of autumn at the InSight landing site: a spectacular burst of daytime vortices, the appearance of nighttime vortices in a flat region, and the development of nighttime local turbulence nearly as intense as during daytime.


\section{InSight's Weather Sensors and Ambient Conditions \label{sec_env}}

InSight's high-frequency high-precision pressure sensor (PS) and the Temperature and Winds for InSight (TWINS) instrument are both part of the Auxiliary Payload Sensor Suite (APSS). 
Furthermore, a radiometer measures the surface brightness temperature \cite{Muel:20}, and 
onboard cameras can be used to quantify atmospheric opacity \cite{Maki:18}. With its complementary and high-sensitivity sensors that worked in continuous mode during more than one year, InSight has provided one of the most complete record ever acquired at the surface of Mars.

In this study, pressure data is chosen to study turbulence (rather than wind data) given the pressure sensor's unprecedented accuracy, high frequency (continuous acquisition in the range 2-10 Hz during the whole mission, compared to 0.1-1 Hz for wind data), and its broader temporal coverage over the mission (e.g. coverage during the 170 first sols of the second martian year is 88\% for pressure versus 33\% for wind).
We applied to all pressure data a low-pass filter with cutoff frequency at 2 Hz, above which mechanical/electrical noise and a loss of effectiveness of the pressure inlet was reported by \citeA{Banf:20}.

\begin{figure}
\centering
\noindent\includegraphics[width=\textwidth]{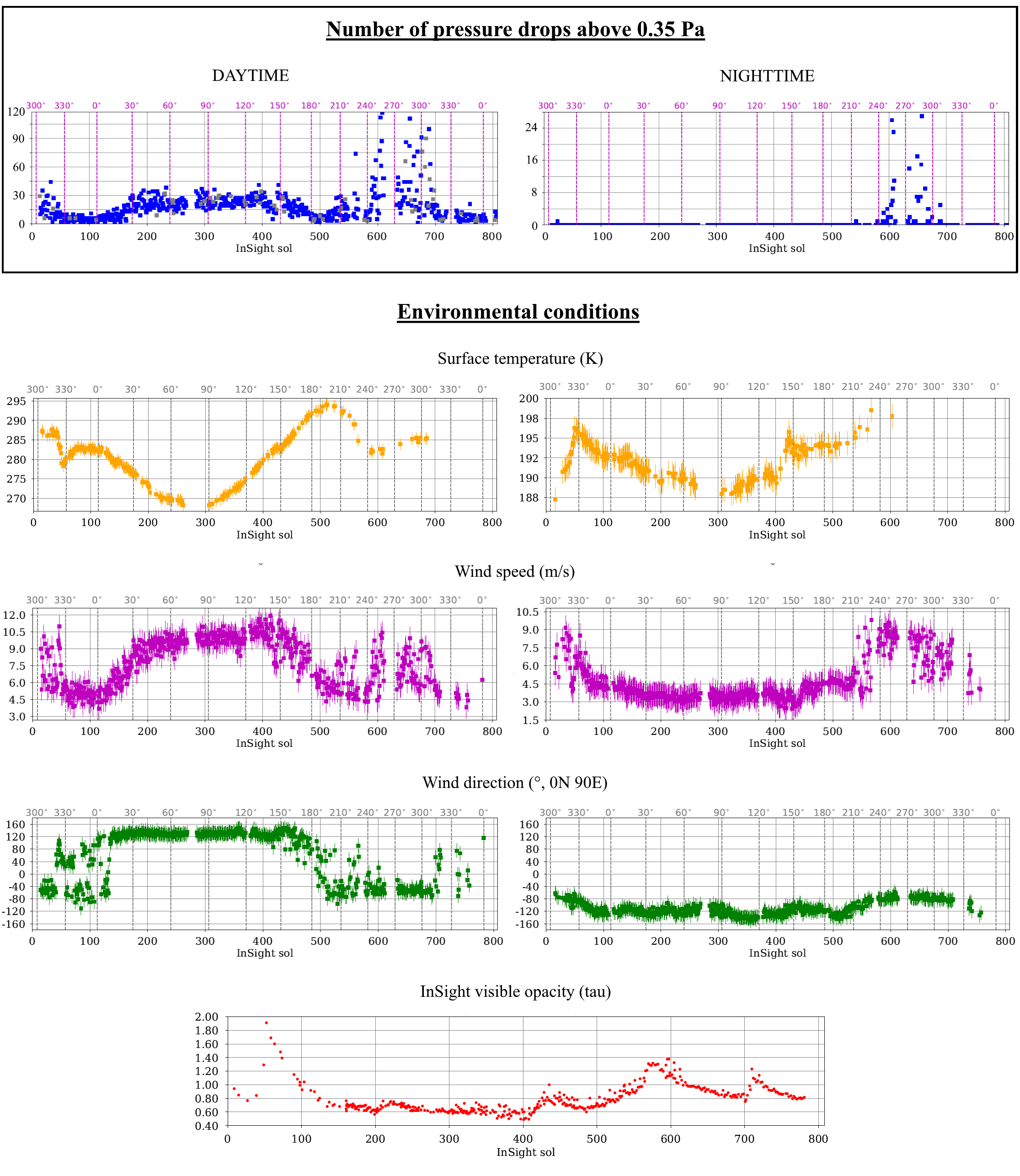}
\caption{Vortex observations and environmental conditions measured by InSight over the first 800 sols of the mission (with one dot per sol). The top panel shows vortex-induced pressure drops above 0.35 Pa amplitude for daytime (left) and nighttime (right), obtained following the detection method described in \citeA{Spig:21jgr} (gray squares correspond to incomplete InSight sols for which the total number of vortices has been reconstructed). The three middle lines show daytime (left, averaged over 12-13 Local True Solar Time LTST) and nighttime conditions (right, 1-3 LTST) with from top to bottom surface temperature, ambient wind speed and ambient wind direction. The last line shows the daytime visible opacity.}
\label{Fig_env}
\end{figure}

To provide context, the InSight observations of environmental variables (surface temperature, dust opacity and wind velocity) are given in Figure~\ref{Fig_env}, extending the plots presented in \citeA{Spig:21jgr} over 800 sols.  
These plots highlight the presence of a very peculiar season from the solar longitudes ($L_{\textrm{s}}$) 210\textdegree~(InSight sol $\sim 550$, northern autumn) to 320\textdegree~(InSight sol $\sim 700$, northern winter). 
At the equatorial InSight landing site, seasons are strongly influenced
by the perihelion ($L_{\textrm{s}}$ 251\textdegree) / aphelion ($L_{\textrm{s}}$ 71\textdegree) variation.
At the $L_{\textrm{s}}$ 210-320\textdegree period, the atmosphere is loaded with dust particles due to seasonal storms, which diminishes the incoming solar flux to the surface and leads to a pronounced decrease of daytime surface temperature. 
Conversely, dust infrared emission induces a moderate increase of the nighttime surface temperature. Furthermore, at this $L_{\textrm{s}} = 210-320^{\circ}$ season, the nighttime ambient wind measured by InSight increases by a factor 2 to 3 compared to the rest of the year, following the strengthening of the low level jet in dustier conditions \cite{Josh:97llj}), while the daytime counterpart undergoes sol-to-sol variability between low values as in northern spring and high values as in northern summer \cite{Spig:21jgr}.
This season also exhibits a change in daytime wind direction, from southeasterlies to northwesterlies (as predicted by models, e.g. \citeA{Spig:18insight}). The nighttime wind direction remains similar all Martian year long.

\section{Vortices \label{sec:vortices}}

\subsection{Method} 
InSight has detected a strong vortex activity since the very beginning of the mission \cite{Banf:20}. 
Previous studies \cite{Spig:21jgr,Lore:20catalog} cataloged daytime vortices observed during the first 400 sols of the mission, corresponding to about half a Martian year, spanning from the end of northern winter to mid-summer. Vortices are detected in the pressure data as sudden pressure drops starting at a threshold of 0.35~Pa.
Following the exact same methodology as in \citeA{Spig:21jgr}, based on search of minima on the pressure data after subtraction of a 1000-second-window moving mean to remove diurnal variations, we extend their vortex catalog to more than a complete Martian year (800 sols) for both day and night.

\subsection{A Burst of Vortices Observed in Autumn/Winter} 

The daily number of vortex-induced pressure drops detected during 1.25 year of InSight mission is reported in Figure~\ref{Fig_env}. We find that a spectacular ``burst'' of daytime 
vortex-induced pressure drops -- with more than a hundred encounters on certain sols, and sol 689 featuring an all-year maximum of 10 encountered vortices deeper than 1 Pa  -- is observed at the InSight site in the northern autumn/winter season (sols 550-700), in stark contrast with the spring/summer season studied in \citeA{Spig:21jgr}.
Compared to other seasons, this high number of vortex encounters 
in autumn/winter is not associated with any seasonal maximum of surface temperature or ambient wind (Figure~\ref{Fig_env}).
Thus, this InSight vortex burst can neither be explained by a stronger PBL forcing by a warmer surface, nor by a more pronounced advection by ambient wind \cite{Spig:21jgr}.

How to explain this daytime vortex burst in northern autumn/winter?
Firstly, since the vortex burst starts when the wind direction changes (section~\ref{sec_env}), 
this opens the possibility that vortices would be advected from a different region with more favourable turbulent conditions; 
however, the surroundings of InSight landing site are extremely uniform with respect to surface properties \cite{Golo:17} hence expected turbulence strength.
Secondly, the vortex burst occurs in the ``dust storm'' season when the atmospheric opacity of the atmosphere is larger than in the aphelion season, but not as large as during a planetary dust event \cite{Guze:19,Mont:20}.
While a dustier atmosphere \emph{a priori} implies lower daytime surface temperature (section~\ref{sec_env}), hence weaker convection and less vortex encounters,
the recent turbulence-resolving simulations of the Martian dusty PBL by \citeA{Wu:21} bears the potential to explain the vortex burst we observe at the dusty season.
A moderately dusty atmosphere may yield a stronger daytime vortex production than a clearer atmosphere, owing to the reinforcement of turbulence by the convergence of dust particles on the walls of convective cells.

The northern autumn/winter season is also prone to a significant number (up to 30 per night, see Figure~\ref{Fig_env}) of abrupt pressure drops during the night, attributed to vortices given the similarities of pressure signatures with their daytime counterparts. 
While daytime vortex-induced pressure drops may reach 10 Pa amplitude \cite{Banf:20,Lore:20catalog}, nighttime pressure drops do not reach more than 1 Pa amplitude.
Nighttime pressure drops are completely absent from other seasons.
The northern autumn/winter season (sols 550-700) is thus uniquely characterized by both daytime and nighttime vortex bursts.

Given the stable inversion layer appearing near the surface on Mars during the night \cite{Mart:17}, the nighttime vortices detected by InSight cannot be caused by buoyancy-driven turbulence as in the daytime.
Interestingly, many nighttime vortex-induced pressure drops have also been observed by Curiosity \cite{Kaha:16,Ordo:18}, tentatively explained by Gale Crater's katabatic winds. 
Conversely, nighttime vortex signatures at the very flat InSight landing site points towards an explanation not requiring the presence of topography.
Stronger nighttime ambient wind conditions in dustier conditions (see section~\ref{sec_env}) is a more plausible explanation and, as we shall see in what follows, the nighttime vortex burst in autumn/winter is part of an overall exceptional nighttime turbulent activity.


\section{Local and Non-Local Turbulence}

\subsection{Method} 

\begin{figure}
\centering
\noindent\includegraphics[width=\textwidth]{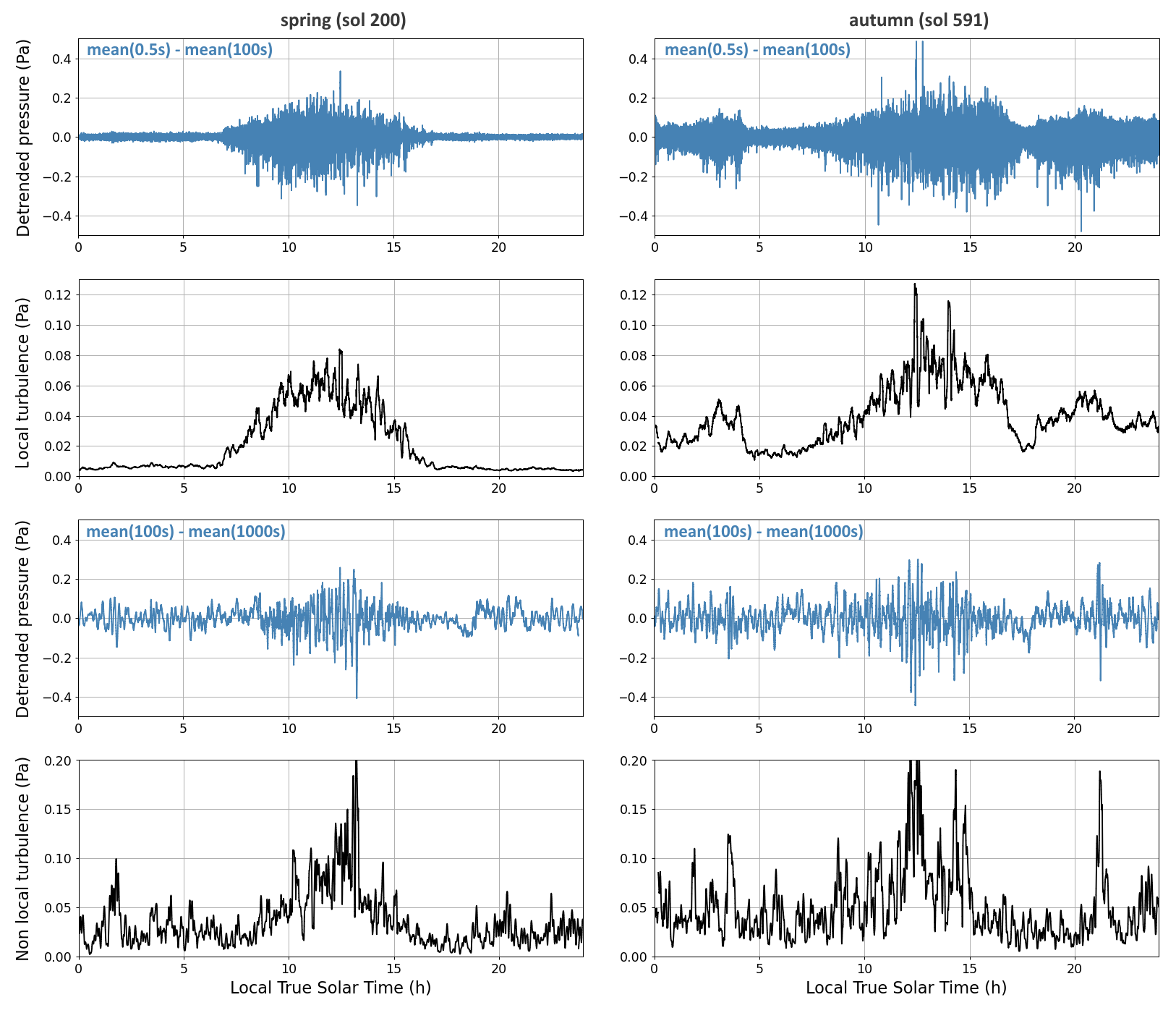}
\caption{Turbulence at InSight landing site for two typical sols: spring ($L_{\textrm{s}}$ = 42\textdegree, left column) and autumn ($L_{\textrm{s}}$ = 246\textdegree, right column). Detrended pressure and standard deviation for local turbulence (top) and non-local turbulence (bottom). Definitions and computations are provided in the main text.}
\label{fig_sol}
\end{figure}

Local turbulence is obtained by detrending InSight pressure time series for each sol with the same time series smoothed by a 100~s window, before computing standard deviation of the result
over a moving window of 500~s.
We also removed from the pressure data the vortex-induced pressure drops detected in section~\ref{sec:vortices}.
Non-local turbulence is obtained by subtracting the 
1000~s-window-averaged pressure to the 100~s-window-averaged pressure, before computing standard deviation of the result over a moving window of 500~s.
To summarize, local turbulence is considered as the range~$0.01-2$~Hz while non-local turbulence the range~$0.002-0.01$~Hz (Figure~\ref{fig_sol}).
Changing the limit between local and non-local turbulence from $0.01$ to $0.02$~Hz do not alter the results (Figure S2).

\subsection{Calm and Active Seasons for Turbulence} 

Figure \ref{fig_sol} and Figure S1 confirm, as is expected from existing observations and numerical modeling \cite{Petr:11,Spig:16ssr,Mart:17}, that non-local turbulence at InSight is mostly significant in daytime -- with a maximum around 12h-13h and with no clear seasonal variations. In nighttime conditions, non-local turbulence is usually weak; the only notable nighttime signal in the non-local frequency range is the occurrence of bores and gravity waves \cite{Banf:20} unrelated to turbulent processes (e.g. sol 591 at 21h in Figure \ref{fig_sol}).

\begin{figure}
\noindent\includegraphics[width=\textwidth]{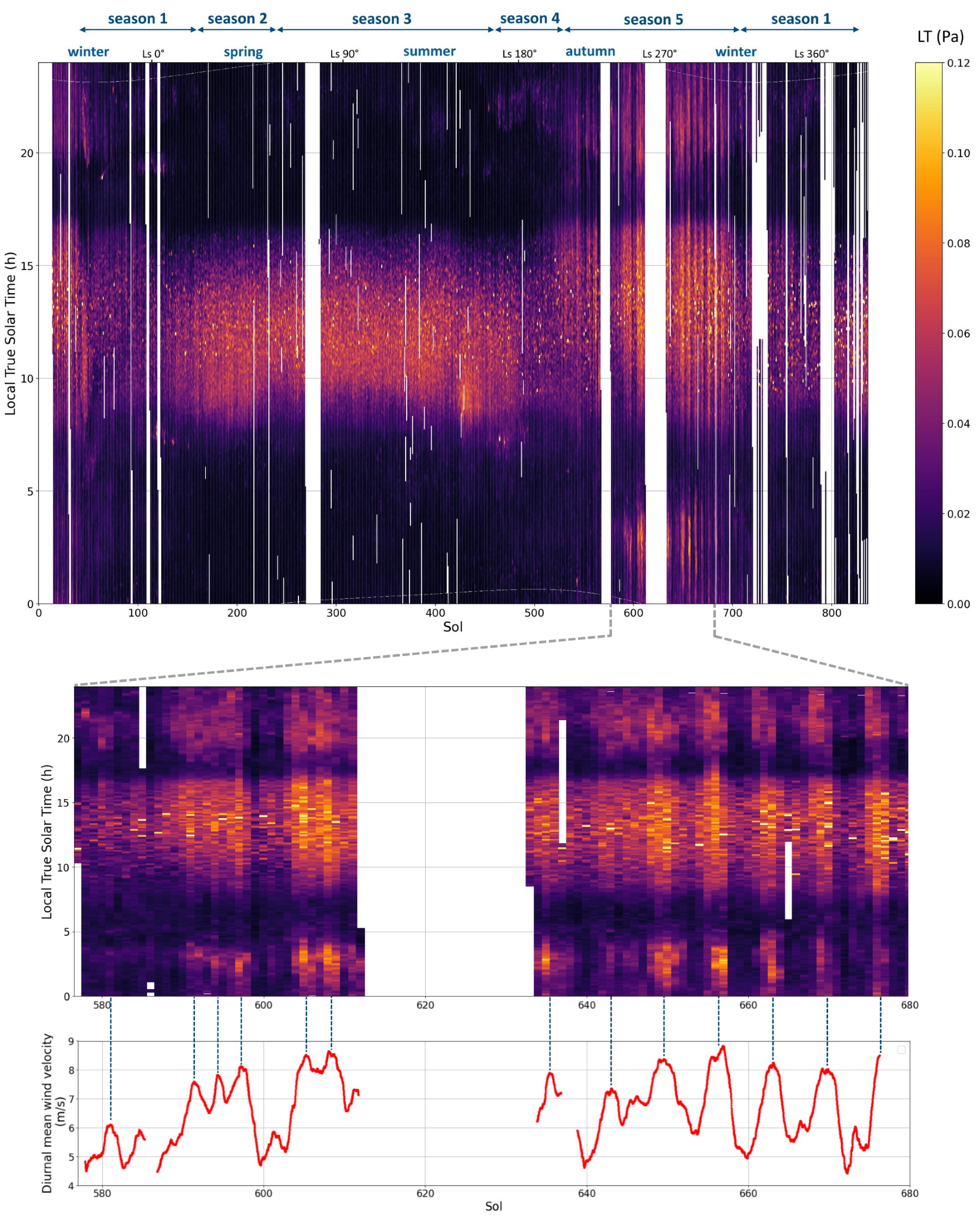}
\caption{Local turbulence during the 837 first sols of the InSight mission, and zoom on the remarkable autumn/winter season, shown with large-scale diurnally-averaged ambient wind (computed with a moving mean at each data point).}
\label{fig_year}
\end{figure}

The remarkable result found in Figure \ref{fig_sol} and best highlighted in Figure \ref{fig_year} is the existence of strong local turbulence at night in the autumn/winter season.
Both in daytime and in nighttime, local turbulence varies strongly with seasons. In Figure \ref{fig_year} we observe that northern spring and summer define a long ``regular'' season (from $L_{\textrm{s}}$ 23\textdegree~to 166\textdegree, later called seasons 2 and 3) characterized by no major sol-to-sol changes of daytime local turbulence and a complete absence of nighttime local turbulence.
This picture changes radically in northern autumn. From $L_{\textrm{s}}=207^{\circ}$ to $L_{\textrm{s}}=320^{\circ}$ (named season 5, at sols 530-710 and 0-40), daytime local turbulence is strong and reaches an annual maximum while, surprisingly so given Mars nights' ultra-stable atmosphere, nighttime local turbulence develops with amplitudes equivalent to daytime local turbulence.
This daytime and nighttime burst of local turbulence at this season echoes the vortex bursts discussed in section~\ref{sec:vortices} and the permanent turbulence-induced seismic noise reported by \citeA{Ceyl:21}.
Interestingly, in the seasons prior and posterior to this remarkable season 5 (seasons 4 and 1 in Figure \ref{fig_year}), a $50\%$ decrease is observed in daytime local turbulence while a small-amplitude nighttime local turbulence is noticeable. 
The availability of InSight data 
for two Martian years at winter (season 1) highlights the strong inter-annual repetition of the seasonal cycle of turbulence.

\subsection{Correlation of Turbulence with Ambient Wind} 

A strong dependence of local turbulence with ambient wind is especially observed during season 5 (autumn/winter) (Figure \ref{fig_year} bottom).
The pseudo-oscillations on periods of 2-3 to 8-10 sols observed on both daytime and nighttime local turbulence are correlated to oscillations in ambient wind, with wind maximum aligned with turbulence maximum. 
These wind variations are due to mid-latitude planetary waves caused by baroclinic instabilities arising from seasonal equator-to-pole temperature gradients \cite{Banf:20}. Such patterns are not observed in non-local turbulence (not shown).

\begin{figure}
\noindent\includegraphics[width=\textwidth]{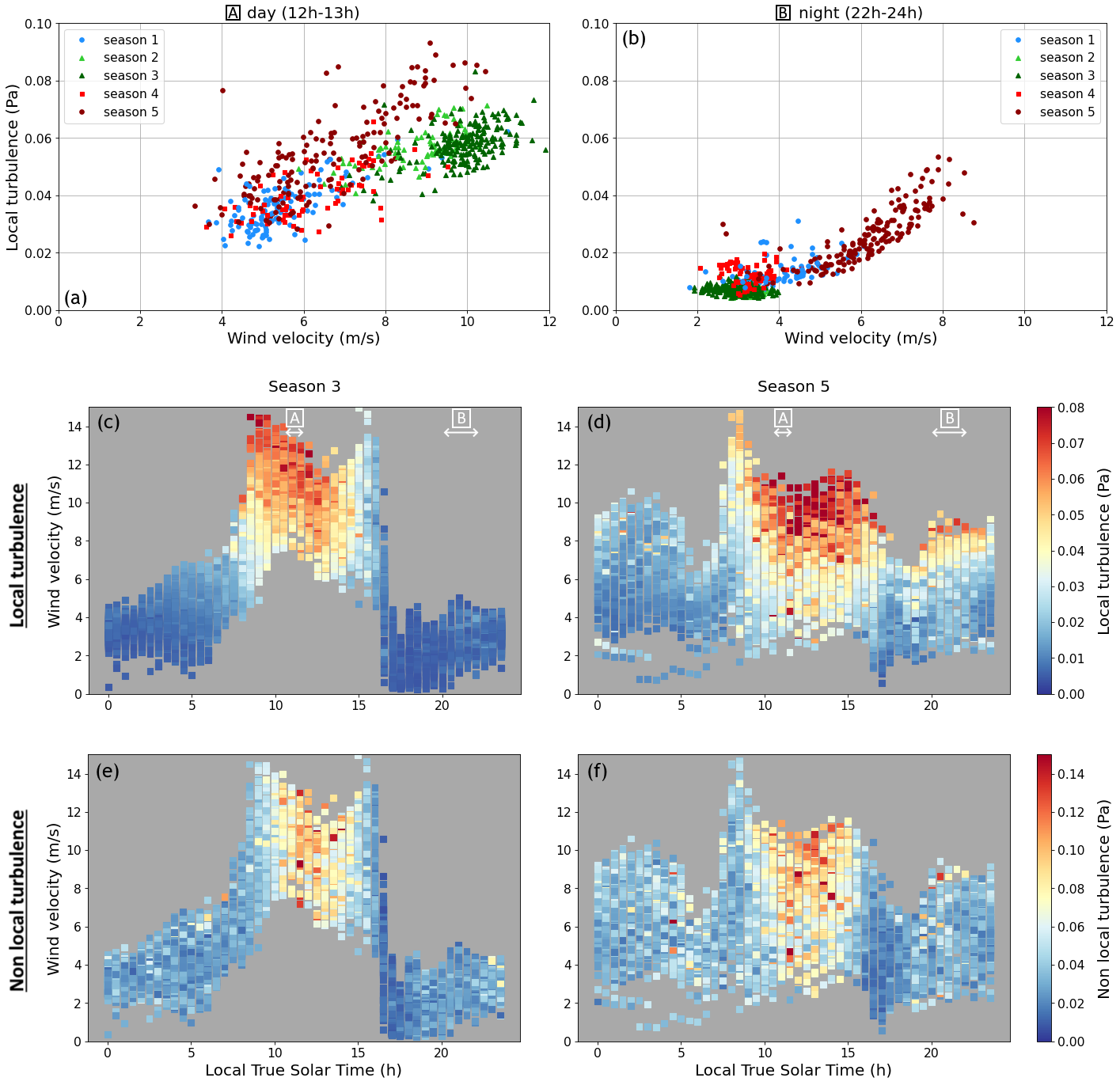}
\caption{Correlation between (a) daytime and (b) nighttime local turbulence and ambient wind for the five seasons identified in Figure~\ref{fig_year}. Diurnal evolution of (c,d) local and (e,f) non-local turbulence with associated ambient wind conditions on the vertical axis; summer (season 3: c,e) and autumn/winter (season 5: d,f).}
\label{fig_wind}
\end{figure}

Figure \ref{fig_wind} (a to d) shows the correlation between ambient wind velocity and local turbulence, and its variation with local time. A clear linear trend is found in daytime, with a higher linear coefficient during the remarkable season 5. The trend found for nighttime observations is clearly different. 
Local turbulence is generally smaller at night compared to the day in the same ambient wind conditions.
Nighttime local turbulence remains weak (standard deviations of 5-15~mPa) for winds lower than a 4~m~s$^{-1}$ velocity threshold, while for larger winds local turbulence increases strongly with ambient wind, with a power law rather than a linear law. This suggests that local turbulence are not driven by the same mechanisms in the daytime vs the nighttime.

In contrast with local turbulence, no noteworthy correlation to wind velocity is observed for non-local turbulence (Figure~\ref{fig_wind}:e,f and Figure S3). This is in line with the conclusions of \citeA{Spig:21jgr} about the non-dependence of wind gustiness to ambient wind velocity, given the lower frequency ($<0.1$~Hz) of InSight wind records only suitable to probe non-local turbulence.

\subsection{Explaining the Nighttime Burst of Local Turbulence in Autumn/Winter}

How could we explain that northern autumn/winter conditions on Mars at the InSight landing site are conducive to strong nighttime turbulence?
Turbulent kinetic energy (TKE) in the PBL takes its origins in two distinct sources: buoyancy and wind shear, since the other terms of the TKE equation simply redistributes TKE \cite{Stul:88,Holt:04}.
In the Martian daytime, the warm surface heated by incoming sunlight is a source of positive buoyancy fueling convective turbulence, through both sensible heat flux and, most importantly, near-surface CO$_2$ absorption of incoming infrared radiation from the surface \cite{Savi:99,Spig:10bl}.
From these daytime buoyancy sources originate both non-local turbulence (plumes, cells) and local turbulence (small eddies).
Shear-driven mechanical turbulence may also act as a source of local turbulence, both in the day and in the night.
In nighttime conditions, near-surface stable inversions act as a turbulence-suppressing buoyancy sink, hence wind shear is the only possible major source of local turbulence; no major source of non-local turbulence is possible at night.

\begin{figure}
\noindent\includegraphics[width=\textwidth]{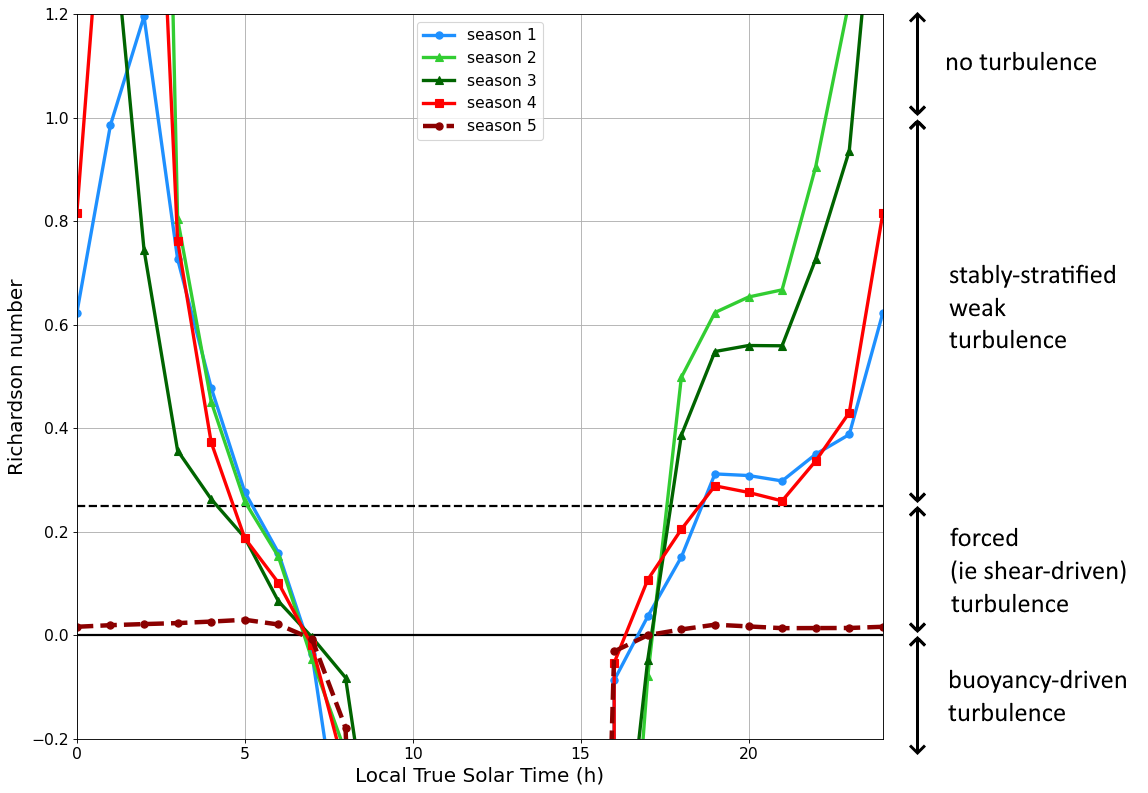}
\caption{Diurnal variations of the Richardson number~$Ri$ computed from the Mars Climate Database at 5 meters above the surface in the conditions of the InSight landing site for the five seasons identified in Figure~\ref{fig_year}. The peculiarity of season 5 (northern autumn/winter) appears clearly.}
\label{fig_Ri}
\end{figure}

A plausible explanation for the nighttime burst of local turbulence experienced by InSight specifically in autumn/winter season is the strong development of the nighttime low-level jet in the dustier conditions at this season, supported both by modeling \cite{Josh:97llj} and InSight ambient observations (section~\ref{sec_env}). However, this nighttime turbulence burst at amplitudes comparable to daytime turbulence (Figures~\ref{fig_sol} and~\ref{fig_year}) remains \emph{a priori} unexpected given the aforementioned strong buoyancy sink for turbulence in Martian nights. 

To investigate further the subtle competition between buoyancy source/sink of turbulence~$\mathcal{B}$ and wind-shear source of turbulence~$\mathcal{S}$, we use the Richardson number~$Ri$ defined as follows (e.g. see \citeA{Stul:88}):
\begin{linenomath*}
\begin{equation}
Ri = \frac{\mathcal{B}}{\mathcal{S}} =  \frac{g}{\theta_0} \frac{ \partial \theta / \partial z}{(\partial V / \partial z)^2}
\end{equation}
\end{linenomath*}
with $z$ height above surface, $g$ gravitational acceleration, $V$ ambient wind velocity, $\theta$ potential temperature with $\theta_0 = \theta(z=0)$. 
Negative~$Ri$ indicates a buoyancy source for convective PBL turbulence in daytime unstable conditions; positive~$Ri$, a buoyancy sink inhibiting turbulence in nighttime stable conditions -- an initially turbulent flow may even become laminar when~$Ri>1$ \cite{Stul:88}. However, an initially laminar flow, e.g. in the strongly stable nighttime conditions on Mars, may become turbulent for $0<Ri<Rc$ where $Rc$ is the critical Richardson number $Rc \simeq 0.25$. In this case, the wind-shear source of turbulence dominates the buoyancy sink (either because the wind shear is strong and/or the stability is weak), enabling nighttime turbulence.
InSight air temperature measurements could not be used to evaluate near-surface temperature gradients for $Ri$ computations, for suspicion of contamination by lander thermal effects \cite{Spig:21jgr,Banf:20}.
Instead, to calculate~$Ri$, we used atmospheric variables 5 meters above surface from the Mars Climate Database \cite{Mill:15} for an average year valid for any non-global-dust-storm Martian year like those covered by InSight.

Computations of the Richardson number $Ri$ for the seasonal conditions experienced by InSight in Figure \ref{fig_Ri} clearly demonstrate that, contrary to other seasons, the northern autumn/winter season is exceptionally propitious to nighttime shear-driven local turbulence, exhibiting weakly positive $Ri$ values in the range $0<Ri<0.25$ all night long.
The dustier atmosphere at this autumn/winter season contributes to decrease the nighttime Richardson number $Ri$ by two means (Figure S4).
Firstly, the strengthening of the nighttime low-level jet deepens the near-surface wind shear ($\mathcal{S}$ increases in~$Ri$).
Secondly, the nighttime increase of surface temperature and decrease of atmospheric temperature, compared to other seasons, weakens near-surface atmospheric stability ($\mathcal{B}$ decreases in~$Ri$).
These two effects in autumn/winter, confirmed by InSight observations of ambient wind and surface temperature in Figure~\ref{Fig_env}, cause the nighttime positive Richardson number to decrease below the critical value~$Rc=0.25$, causing the nighttime burst of local turbulence unveiled by Figure~\ref{fig_year} and the associated formation of nighttime vortices reported in Figure~\ref{Fig_env}.

We note that seasons 4 and 1 which precedes and follows the remarkable autumn/winter season exhibit evening conditions slightly above the $Ri$ limit of 0.25, which might explain why those seasons show moderate nighttime turbulence (Figure~\ref{fig_year}).
Conversely, the nighttime conditions experienced in northern spring/summer at InSight (seasons 2 and 3) are clearly in the $Ri$ domain where buoyancy sink dominates over shear source, hence the absence of observed nighttime local turbulence (Figure~\ref{fig_year}).


\section{Conclusion}

The InSight mission brought an unprecedented complete dataset to study atmospheric turbulence at the surface of Mars and variability thereof at the seasonal timescales. In this work, we based our analysis on high-frequency high-accuracy continuous pressure data to investigate vortices (as pressure drops), local turbulence (phenomena shorter than 100~s) and non-local turbulence due to larger scale structures such as convection cells and plumes (inducing signals of 100-500~s).
Contrary to non-local turbulence, local turbulence inferred from InSight data is strongly sensitive at all local times and seasons to ambient wind.

The autumn/winter ``dusty'' season around the perihelion ($L_{\textrm{s}}$ 251\textdegree) at the InSight landing site is a remarkably active season for turbulence.
\begin{itemize}
\item In daytime, InSight experiences a burst of vortex encounters that is yet to be explained, for it does not correspond to any significant change in environmental conditions more favorable to a rise in vortex encounters, though the feedback of transported dust particles on PBL convection is a possible explanation worth being investigated further \cite{Wu:21}. Moreover, this InSight vortex burst makes an interesting challenge for predictive diagnostics such as dust devil activity \cite{Newm:19}.
\item In nighttime, unexpectedly so given ultra-stable conditions, shear-driven local turbulence at the InSight landing site is almost as powerful as in the convective daytime, as a result of both the stronger ambient wind (low-level jet) and the weaker stability in dustier conditions at this season.
Vortices akin to those appearing in daytime conditions (albeit of weaker amplitude) appear only at this season at the InSight site.
\end{itemize}

Our results demonstrate that the continuous high-frequency atmospheric measurements by InSight are key to unveil the properties of Martian turbulence.  
Emphasis for future modeling work could be on reproducing and explaining, at the remarkable autumn/winter season unveiled by InSight, both the burst of daytime vortices and the exceptionally high level of nighttime turbulence -- the latter being a difficult challenge for terrestrial models \cite{mahr:98stratified}.
Furthermore, our conclusions on seasonal bursts of turbulence, especially at night, may have key implications for dust storm dynamics and aeolian processes.



\acknowledgments

This is InSight contribution number 232. The authors acknowledge the funding support provided by Agence Nationale de la Recherche (ANR-19-CE31-0008-08 MAGIS) and CNES. All co-authors acknowledge NASA, Centre National d’Études Spatiales (CNES) and its partner agencies and institutions (UKSA, SSO, DLR, JPL, IPGP-CNRS, ETHZ, IC, and MPS-MPG), and the flight operations team at JPL, CAB, SISMOC, MSDS, IRIS-DMC, and PDS for providing InSight data. All InSight data used in this study are publicly available in the Planetary Data System. Data from the APSS pressure sensor and the temperature and wind (TWINS) sensor referenced in this study is publicly available from the PDS (Planetary Data System) Atmospheres node. The direct link to the InSight data archive at the PDS Atmospheres node is: \url{https://atmos.nmsu.edu/data_and_services/atmospheres_data/INSIGHT/insight.html}. Surface brightness temperature measured by the HP$^3$ radiometer is publicly available from the PDS Geosciences node: \url{https://pds-geosciences.wustl.edu/missions/insight/hp3rad.htm} (DOI reference \url{https://doi.org/10.17189/1517568}). The datasets produced to obtain the figures in this study, along with the Python codes used for the data analysis, are available in the citable online archive \url{https://doi.org/10.14768/07197931-C0D4-41B1-A437-DE516D0086F3}.


\bibliography{Turbulence_InSight}

\end{document}